\documentclass[a4paper,11pt]{article}
\usepackage{pos}

\usepackage{multirow} 
\usepackage{subcaption}

\usepackage{physics}
\usepackage{enumitem}
\usepackage[htt]{hyphenat}
\usepackage{soul} 

\usepackage{colortbl}
\definecolor{lightblue}{RGB}{202, 225, 255}
\definecolor{olivegreen}{RGB}{202, 255, 112}
\definecolor{darkolivegreen}{RGB}{85, 107, 47}
\definecolor{littledarkgreen}{RGB}{0,80,0}
\definecolor{firebrick}{RGB}{178,34,34}
\definecolor{darkslateblue}{RGB}{72,61,139}
\definecolor{midnightblue}{RGB}{25,25,112}
\definecolor{darkblue}{RGB}{0,0,139}
\definecolor{indigo}{RGB}{75,0,130}
\definecolor{dodgerblue}{RGB}{30,144,255}
\definecolor{mistyrose}{RGB}{255,228,225}
\definecolor{khaki}{RGB}{240,230,140}
\definecolor{royalblue}{RGB}{65,105,255}
\definecolor{mediumseagreen}{RGB}{60,179,113}
\definecolor{mediumspringgreen}{RGB}{60,179,113}
\definecolor{lime}{RGB}{0,255,0}
\definecolor{limegreen}{RGB}{50,205,50}
\definecolor{teal}{RGB}{0,128,128}
\definecolor{blueviolet}{RGB}{138,43,226}
\definecolor{lightmagenta}{RGB}{255,51,255}
\definecolor{background1}{RGB}{245,255,250}

\providecommand{\fm}{\;\mathrm{fm}}

\graphicspath{{./Figs/}}

\title{Form factors for semileptonic B-decays with HISQ light quarks and clover b-quarks in Fermilab interpretation}
\ShortTitle{Semileptonic B-decays with HISQ light quarks and Fermilab(clover) b-quarks}

\author*[a]{Hwancheol Jeong}
\author[b]{Carleton DeTar}
\author[c]{Aida El-Khadra}
\author[d]{Elvira G\'amiz}
\author[c]{Zechariah Gelzer}
\author[a]{Steven Gottlieb}
\author[e]{William Jay}
\author[f]{Andreas Kronfeld}
\author[c]{Andrew Lytle}
\author[g]{Alejandro Vaquero}

\affiliation[a]{Department of Physics, Indiana University, Bloomington, Indiana 47405, USA}
\affiliation[b]{Department of Physics and Astronomy, University of Utah, Salt Lake City, Utah 84112, USA}
\affiliation[c]{Department of Physics, University of Illinois, IL 61801, USA}
\affiliation[d]{CAFPE and Departamento de F\'isica Te\'orica y del Cosmos, Universidad de Granada, E-18071, Granada, Spain}
\affiliation[e]{Center for Theoretical Physics, Massachusetts Institute of Technology, Cambridge, MA 02139}
\affiliation[f]{Theory DivisionFermi National Accelerator Laboratory, Batavia, Illinois, USA}
\affiliation[g]{CAPA \& Departamento de F\'isica Te\'orica, Universidad de Zaragoza, 50009 Zaragoza, Spain}

\emailAdd{sonchac@gmail.com}

\abstract{
  We compute the vector, scalar, and tensor form factors for the $B\to \pi$, $B\to K$, and $B_s\to K$ amplitudes, which are needed to describe semileptonic $B$-meson decay rates for both the charged and neutral current cases. We use the highly improved staggered quark (HISQ) action for the sea and light valence quarks. The bottom quark is described by the clover action in the Fermilab interpretation. Simulations are carried out on $N_f = 2+1+1$ MILC HISQ ensembles at approximate lattice spacings from $0.15 \fm$ down to $0.057 \fm$. We present blinded preliminary results for the form factors.
}

\FullConference{The 40th International Symposium on Lattice Field Theory (Lattice 2023)\\
July 31st - August 4th, 2023\\
Fermi National Accelerator Laboratory\\}

\begin{document}
\maketitle

\section{Introduction}
\label{sec:intro}
This work is an update on ongoing lattice QCD calculation of form factors for the $B\to \pi$, $B\to K$, and $B_s\to K$ amplitudes \cite{Gelzer:2017edb,FermilabLattice:2019qpy}. We use the highly improved staggered quark (HISQ) action~\cite{Follana:2006rc} for the sea and light valence quarks. The bottom quark is described by the clover action in the Fermilab interpretation~\cite{El-Khadra:1996wdx}. Simulations are carried out on (2+1+1)-flavor MILC HISQ ensembles \cite{MILC:2012znn}. Continuing from Refs.~\cite{Gelzer:2017edb,FermilabLattice:2019qpy}, we have changed some of our data analysis strategies and redid the analysis. We present some preliminary results for the lattice form factors.

In Sec.~\ref{sec:ff}, we briefly describe the form factors of interest. 
In Sec.~\ref{sec:lattice}, the simulation details and our data analysis strategy are described. In Sec.~\ref{sec:anal_two}, we analyze our two-point correlation function data and extract mesons' ground and a few excited eigenstate information. In Sec.~\ref{sec:anal_ff}, we calculate the lattice form factors from the two- and three-point correlation function data and present some preliminary results. Summary and future plans are discussed in Sec.~\ref{sec:conc}.

\section{Form factors}
\label{sec:ff}
Transitions between pseudoscalar mesons are described by the matrix elements of a vector current $\mathcal{V}^\mu = \overline{q} \gamma^\mu b$, a tensor current $\mathcal{T}^{\mu \nu} = i\, \overline{q} \sigma^{\mu \nu} b$, and a scalar current $S = \overline{q} b$, that can be expressed in terms of the form factors $f_{+}$, $f_0$, and $f_T$. The details are described in Ref.~\cite{Gelzer:2017edb}. For convenience, we consider the following form factors
\begin{align}
  f_\parallel (E_L) &= \frac{ \mel{L}{\mathcal{V}^0}{B} }{ \sqrt{2 M_B} } \,,\\
  f_\perp (E_L) &= \frac{ \mel{L}{\mathcal{V}^i}{B} }{ \sqrt{2 M_B} } \frac{1}{k^i} \,,\\
  f_T (E_L) &= \frac{M_B + M_L}{ \sqrt{2 M_B} } \frac{ \mel{L}{\mathcal{T}^{0i}}{B} }{ \sqrt{2 M_B} } \frac{1}{k^i} \,,
\end{align}
where $B = B, B_{s}$ represents the $B_{(s)}$ mesons in the initial state, $L = \pi, K$ represents the light pseudoscalar mesons in the final state, $M_{B(L)}$ is the $B(L)$ meson mass, $E_L$ is the $L$ meson recoil energy in the $B$ meson rest frame, and $k$ is the $L$ meson four-momentum. We obtain $f_\parallel$, $f_\perp$, and $f_T$ by analyzing two- and three-point correlation functions. $f_{+}$ and $f_0$ can be constructed from linear combinations of form factors $f_\parallel$ and $f_\perp$.

\section{Lattice calculation}
\label{sec:lattice}
Our lattice calculations are carried out on the (2+1+1)-flavor gauge configurations generated by the MILC Collaboration \cite{MILC:2012znn}. The sea and light valence quarks are simulated with the HISQ action. For the bottom quarks, we use the Sheikholeslami-Wohlert (SW) clover action in the Fermilab interpretation. We employ seven gauge ensembles at approximate lattice spacings $a$ from $0.15$ fm down to $0.057$ fm. Details about these ensembles and our simulation setup are explained in Ref.~\cite{FermilabLattice:2019qpy}.

We measure two- and three-point correlation functions $C^{B(L)}_2$ and $C^{B \to L}_3$ for $B=B_{(s)}$ and $L=\pi,K$ as described in Ref.~\cite{Gelzer:2017edb}. The interpolating operators for the $L$ meson two-point correlation functions and the three-point correlation functions are not smeared. Meanwhile, the interpolating operators for the $B$ meson two-point correlation functions are smeared by the Richardson 1S wave function at both the sources and the sinks \cite{Richardson:1978bt}. The $L$ meson momenta are generated up to $\vectorbold{k} \equiv \vectorbold{n}_{\vectorbold{k}} \times 2\pi / (aN_s) = (4,0,0) \times 2\pi / (aN_s)$. Using the spectral decomposition, we can write them as
\begin{align}
  C^B_2 (t;\vectorbold{0}) &= \sum_{n=0}^{2N-1} (-1)^{n(t+1)} \frac{\left| Z_B^{(n)} \right|^2}{2M_B^{(n)}} \left[ e^{-M_B^{(n)}t} + e^{-M_B^{(n)} (N_t - t) } \right] \,, \label{eq:corr2B} \\
  C^L_2 (t;\vectorbold{k}) &= \sum_{n=0}^{2N-1} (-1)^{n(t+1)} \frac{\left| Z_L^{(n)} \right|^2}{2E_L^{(n)}} \left[ e^{-E_L^{(n)}t} + e^{-E_L^{(n)} (N_t - t) } \right] \,, \label{eq:corr2L} \\
  [C_3^{B \to L}]^{\mu(\nu)} (t,T;\vectorbold{k}) &= \sum_{m,n=0}^{2N-1} (-1)^{m(t+1)} (-1)^{n(T-t-1)} \frac{Z_L^{(m)}}{2 E_L^{(m)}} A_{mn}^{\mu(\nu)} \frac{Z_B^{\dagger (n)}}{2 M_B^{(n)}} e^{-E_L^{(m)}t} e^{-M_B^{(n)} (T - t) } \,, \label{eq:corr3} 
\end{align}
where $Z_B^{(n)} = \mel{0}{O_B}{B^{(n)}}$, $Z_L^{(n)} = \mel{0}{O_L}{L^{(n)}}$, and $A_{mn}^{\mu(\nu)} = \mel{L^{(m)}}{J^{\mu(\nu)}}{B^{(n)}}$ for the lattice currents $J^{\mu(\nu)} = V^\mu$, $T^{\mu\nu}$. We also consider a ratio of correlation functions \cite{Bailey:2008wp}
\begin{equation}
  R(t,T) = \frac{ C_3^{B \rightarrow L}(t,T) }{ \sqrt{C_2^L(t) C_2^B(T-t)} } \sqrt{ \frac{ 2 E_L^{(0)} }{ e^{-E_L^{(0)}t} e^{-M_B^{(0)}(T-t)} } } \,. \label{eq:ratio}
\end{equation}
Inserting Eqs.~\eqref{eq:corr2B} and \eqref{eq:corr2L} into Eq.~\eqref{eq:ratio}, $R$ can be expressed as
\begin{align}
  R(t,T) \simeq \frac{A_{00}}{\sqrt{2 M_B^{(0)}}} \Bigg[ 1 &+ \sum_{m=1} (-1)^{m(t+1)} \left( \frac{\widetilde{A}_{m0}}{\widetilde{A}_{00}} - \frac{1}{2} \left( \frac{\widetilde{Z}_L^{(m)}}{\widetilde{Z}_L^{(0)}} \right)^2 \right) e^{- \delta E_L^{(m)} t} \nonumber\\
    &+ \sum_{n=1} (-1)^{n(T-t-1)} \left( \frac{\widetilde{A}_{0n}}{\widetilde{A}_{00}} - \frac{1}{2} \left( \frac{\widetilde{Z}_B^{(n)}}{\widetilde{Z}_B^{(0)}} \right)^2 \right) e^{- \delta M_B^{(n)} (T-t)} \nonumber\\
    &+ \sum_{m,n=1} (-1)^{m(t+1)} (-1)^{n(T-t-1)} \left( \frac{\widetilde{A}_{mn}}{\widetilde{A}_{00}} + \frac{1}{4} \left( \frac{ \widetilde{Z}_L^{(m)} \widetilde{Z}_B^{(n)} }{ \widetilde{Z}_L^{(0)} \widetilde{Z}_B^{(0)} } \right)^2 \right) e^{- \delta E_L^{(m)} t} e^{- \delta M_B^{(n)} (T-t)} 
    + (\cdots) \Bigg] \,, \label{eq:ratio2}
\end{align}
where $\displaystyle \widetilde{Z}_B^{(n)} \equiv \frac{\left| Z_B^{(n)} \right|}{\sqrt{2M_B^{(n)}}}$, $\displaystyle \widetilde{Z}_L^{(n)} \equiv \frac{\left| Z_L^{(n)} \right|}{\sqrt{2E_L^{(n)}}}$, $\displaystyle \widetilde{A}_{mn} = \frac{Z_L^{(m)}}{2 E_L^{(m)}} A_{mn} \frac{Z_B^{\dagger (n)}}{2 M_B^{(n)}}$, $\delta E_L^{(m)} = E_L^{(m)} - E_L^{(0)}$, and $\delta M_B^{(n)} = M_B^{(n)} - M_B^{(0)}$. The omitted terms are composed of three or four exponential factors.

We obtain $M_B^{(n)}$, $E_L^{(n)}$, and $Z_{B(L)}^{(n)}$ by fitting the two-point correlation functions to Eqs.~\eqref{eq:corr2B} and \eqref{eq:corr2L}. The fitted ground state energy (or mass) is used in computing the ratio $R$ as in Eq.~\eqref{eq:ratio}. The excited state energies are used as priors for fitting the ratio. We extract the ground state matrix elements $A_{00}^{\mu(\nu)}$ and the corresponding form factors from the leading constant term of $R$.

\section{Data analysis: two-point correlation function}
\label{sec:anal_two}
\begin{figure}[tb]
  \centering
  \includegraphics[width=0.8\linewidth]{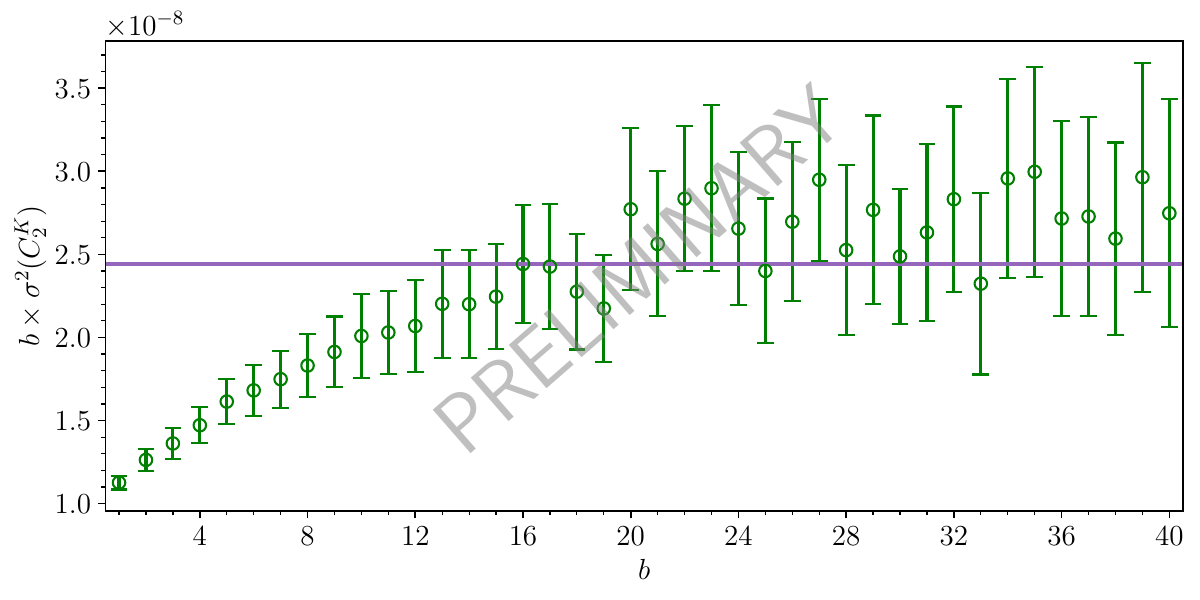}
  \caption{Variance of the binned kaon two-point correlation function $C^K_2 (t=t_{\textrm{min}})$ with increasing bin size $b$. Here, $a \simeq 0.088 \fm$, $m_l / m_s = \textrm{physical}$, and $\vectorbold{n}_{\vectorbold{k}} = (1,1,0)$. The purple line represents the value at the chosen bin size $b=16$.}
  \label{fig:binsize}
\end{figure}
We apply binning on our data to mitigate the autocorrelation between successive configurations. The bin sizes are chosen by monitoring the variance of the two-point correlation function over various bin sizes. In Fig.~\ref{fig:binsize}, we show the variance of kaon two-point correlation function $C^K_2(t)$ at a time slice $t=t_{\textrm{min}}$ as the bin size $b$ increases. When multiplied by $b$, the saturation of increase implies that the remaining autocorrelation is negligible. In the example in the figure, we choose a bin size of $16$.
We also measure the autocorrelation function with the chosen bin size and check whether it is statistically negligible.
In this way, we choose a reasonable minimum bin size for the two-point function data for each meson ($B$, $B_s$, $\pi$, and $K$) for each momentum $\vectorbold{k}$. We then use the largest bin size among this set for all data in a given ensemble.

For a two-point correlation function $C_2(t)$, we define the effective mass and the effective amplitude as
\begin{align}
  a M_{\textrm{eff}} (t) &= \textrm{cosh}^{-1} \left( \frac{C_2(t+1) + C_2(t-1)}{2\, C_2(t)} \right) \,, \label{eq:meff} \\
  A_{\textrm{eff}} (t) &= \frac{C_2(t)}{ e^{-M_{\textrm{eff}}\,t} + e^{-M_{\textrm{eff}}(N_t-t)} } \,. \label{eq:aeff}
\end{align}
However, correlation functions for staggered fermions have both positive and negative parity contributions, where the former oscillates in Euclidean time. This oscillating contribution is significant for heavy mesons, i.e., $B_{(s)}$ mesons.
To suppress the oscillating contribution and obtain a better estimate of the effective mass and amplitude, we consider an averaged (or smeared) two-point correlation function $\overline{C}_2$ defined as \cite{Bailey:2008wp}
\begin{equation}
  \overline{C}_2 (t) \equiv \frac{ e^{-E^{(0)}t} }{4} \left[ \frac{C_2(t)}{e^{-E^{(0)}t}} + \frac{2 C_2(t+1)}{e^{-E^{(0)}(t+1)}} + \frac{C_2(t+2)}{e^{-E^{(0)}(t+2)}} \right]  \,, \label{eq:c2_avg}
\end{equation}
where $E^{(n)}$ represents $E_L^{(n)}$ or $M_B^{(n)}$. It suppresses the oscillating contribution by a factor of $(E^{(1)}-E^{(0)}) / 4$ \cite{Bailey:2008wp}. However, since the ground state energy $E^{(0)}$ has not been obtained at this point, we compute $M_{\textrm{eff}}$ using Eq.~\eqref{eq:meff} and substitute it for $E^{(0)}$. Then the effective mass and amplitude are computed again using Eq.~\eqref{eq:meff} and \eqref{eq:aeff} with $C_2$ replaced by $\overline{C}_2$.
We find that the averaging method gives a more precise estimate for the fitted ground state mass.
We compute the effective mass and amplitude with the averaging method and use them as priors for the ground state mass and amplitude with some relaxed prior widths when we fit the two-point correlation function data.

\begin{figure}[tb]
  \centering
  \includegraphics[width=0.8\linewidth]{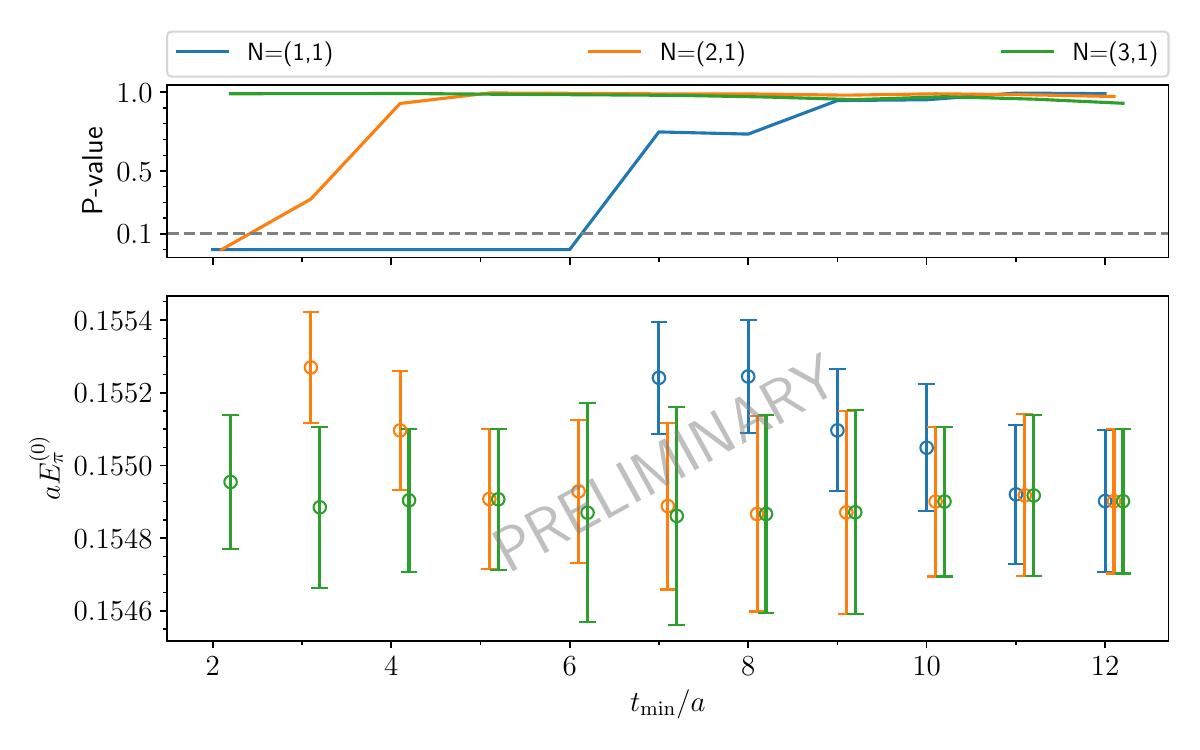}
  \caption{Stability test of the ground state energy of the pion with $\vectorbold{n}_{\vectorbold{k}} = (1,0,0)$ over various $t_{\textrm{min}}$'s and $N$'s. Here, $a \simeq 0.12 \fm$ and $m_l / m_s = \textrm{physical}$. P-values are computed from $\chi^2$ with the augmented terms (due to the Bayesian priors) removed.}
  \label{fig:tmin}
\end{figure}
We fit two-point correlation function data to the functional forms Eqs.~\eqref{eq:corr2B} and \eqref{eq:corr2L} with given $N \equiv ( N_{\textrm{no}}, N_{\textrm{o}} )$ for the number of oscillating (non-oscillating) states $N_{\textrm{o(no)}}$. Fit ranges $[ t_{\textrm{min}}, t_{\textrm{max}} ]$ are chosen for $N = (1,1)$, $(2,1)$, and $(3,1)$ so that their fit posteriors (primarily the ground state energy) are consistent as well as stable under slight variations of $t_{\textrm{min}}$ and $t_{\textrm{max}}$. $t_{\textrm{max}}$ is also set so that the errors of the included $C_2(t)$ data are less than 5\%. Figure \ref{fig:tmin} shows an example of how we determine reasonable $t_{\textrm{min}}$'s. We perform the two-point function fitting for each $N$ while varying $t_{\textrm{min}}$. Examining the distribution of the fitted ground state energies, we may choose $t_{\textrm{min}}/a = 11$ for $N = (1,1)$, $t_{\textrm{min}}/a = 5$ for $N = (2,1)$, and $t_{\textrm{min}}/a = 2$ for $N = (3,1)$. The energies are consistent for a few larger $t_{\textrm{min}}/a$'s and also are consistent across $N$'s. We take the fit posteriors for $N = (2,1)$ in our analysis.

\begin{figure}[tb]
  \centering
  \begin{subfigure}[t]{0.495\linewidth}
    \includegraphics[width=\linewidth]{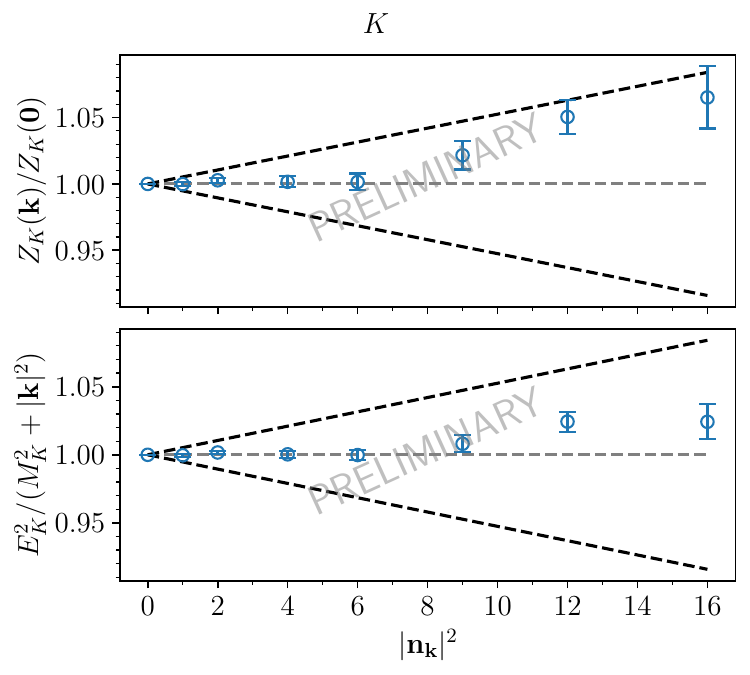}
    \caption{}
    \label{fig:disp_K}
  \end{subfigure}
  \hfill
  \begin{subfigure}[t]{0.495\linewidth}
    \includegraphics[width=\linewidth]{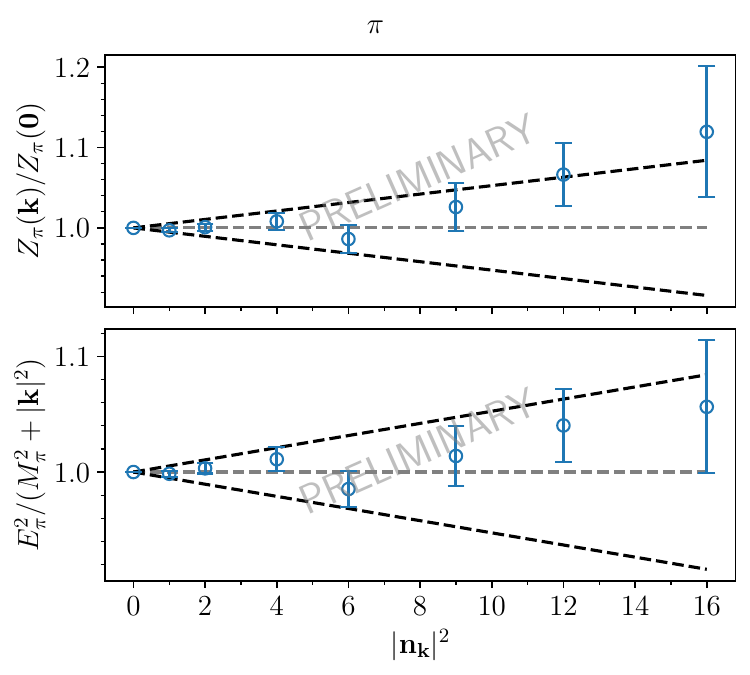}
    \caption{}
    \label{fig:disp_pi}
  \end{subfigure}
  \caption{Comparison of the ground state energy (bottom) and amplitude (top) by direct measure and those by dispersion relation. Here, $a \simeq 0.12 \fm$, and $m_l / m_s = \textrm{physical}$.}
  \label{fig:disp}
\end{figure}
In Fig.~\ref{fig:disp}, we test the dispersion relation $E^2 = M^2 + \vectorbold{k}^2$ for the fitted ground state energies $E^{(0)}_L (\vectorbold{k})$ and the consistency of the corresponding amplitudes $Z_L (\vectorbold{k})$ compared with their zero momentum values $E^{(0)}_L (\vectorbold{0}) = M^{(0)}_L$ and $Z_L (\vectorbold{0})$, respectively. The dashed lines indicate the discretization errors $\mathcal{O}(\alpha_s a^2 \vectorbold{k}^2 )$ from a power counting estimate. The results show that the ratios lie within the dashed cones, within the statistical uncertainty. However, since the signal-to-noise ratio for the correlation function decreases as the momentum increases, the fit posteriors tend to have bigger uncertainties for larger momenta. Hence, $E^{(0)}_L (\vectorbold{k})$ computed from the dispersion relation with the ground state mass $M^{(0)}_L$ at $\vectorbold{k}=\vectorbold{0}$ has a smaller error than those directly measured. We take the values obtained from the dispersion relation in our analysis.

\section{Data analysis: form factor}
\label{sec:anal_ff}
We define an averaged (or smeared) three-point correlation function $\overline{C}_3$ \cite{Bailey:2008wp}:
\begin{align}
  \overline{C}^{B\to L}_3 &(t,T) \equiv \frac{ e^{-E_L^{(0)}t} e^{-M_B^{(0)}(T-t)} }{8} 
  \Bigg[
    \frac{ C^{B \to L}_3 (t,T) }{ e^{-E_L^{(0)}t} e^{-M_B^{(0)}(T-t)} } +
    \frac{ C^{B \to L}_3 (t,T+1) }{ e^{-E_L^{(0)}t} e^{-M_B^{(0)}(T+1-t)} } +
    \frac{ 2 C^{B \to L}_3 (t+1,T) }{ e^{-E_L^{(0)}(t+1)} e^{-M_B^{(0)}(T-t-1)} } \nonumber\\ &+
    \frac{ 2 C^{B \to L}_3 (t+1,T+1) }{ e^{-E_L^{(0)}(t+1)} e^{-M_B^{(0)}(T-t)} } +
    \frac{ C^{B \to L}_3 (t+2,T) }{ e^{-E_L^{(0)}(t+2)} e^{-M_B^{(0)}(T-t-2)} } +
    \frac{ C^{B \to L}_3 (t+2,T+1) }{ e^{-E_L^{(0)}(t+2)} e^{-M_B^{(0)}(T-t-1)} } \Bigg] \,,
\end{align}
which suppresses the oscillating states' contribution in a manner similar to $\overline{C}_2$ defined in Eq.~\eqref{eq:c2_avg} \cite{Bailey:2008wp}.
We compute the averaged ratio $\overline{R}(t,T)$ as defined in Eq.~\eqref{eq:ratio} but with the averaged correlation functions $\overline{C}_2$ and $\overline{C}_3$ in place of $C_2$ and $C_3$, respectively.
Referring to Eq.~\eqref{eq:ratio2}, we have tried various fit models for $\overline{R}(t,T)$ and found that the following fit model describes our data well:
\begin{alignat}{2}
\overline{R} (t,T) \sim F^{(0)} \Big[ 1 \  &+ (-1)^{t+1} F_L^{(1)} e^{- \delta E_L^{(2)} t} &&+ F_L^{(2)} e^{- \delta E_L^{(2)} t} \nonumber\\
  &+ (-1)^{T-t-1} F_B^{(1)} e^{- \delta M_B^{(2)} (T-t)} &&+ F_B^{(2)} e^{- \delta M_B^{(2)} (T-t)} \Big] \,, \label{eq:fitModel}
\end{alignat}
which is composed of the first excited oscillating and non-oscillating states' contributions from both mesons. Here, $F_{(L,B)}^{(n)}$, $\delta E_L^{(n)}$, and $\delta M_B^{(n)}$ are fit parameters. For the latter two, we use the two-point function analysis results as priors with some relaxed uncertainties. We perform the fitting of our averaged ratio data with this fit model. Exceptions are that for $B \to \pi$ and $B \to K$ decays, the oscillating contribution from the $L$ meson is excluded for better goodness-of-fit.

\begin{figure}[tb]
  \captionsetup[subfigure]{aboveskip=0.em,belowskip=0.5em}
  \centering
  \begin{subfigure}[t]{\linewidth}
    \includegraphics[width=\linewidth]{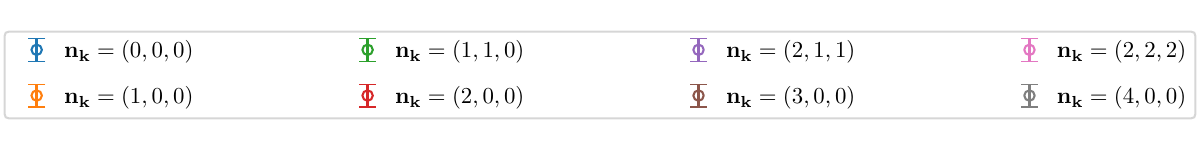}
  \end{subfigure}
  
  \begin{subfigure}[t]{.495\linewidth}
    \includegraphics[width=\linewidth]{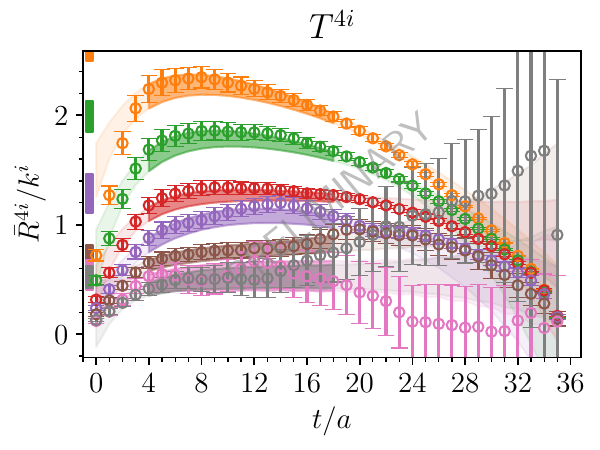}
    \caption{$B \to \pi$}
    \label{fig:ratio_B2pi_T14}
  \end{subfigure}
  \hfill
  \begin{subfigure}[t]{.495\linewidth}
    \includegraphics[width=\linewidth]{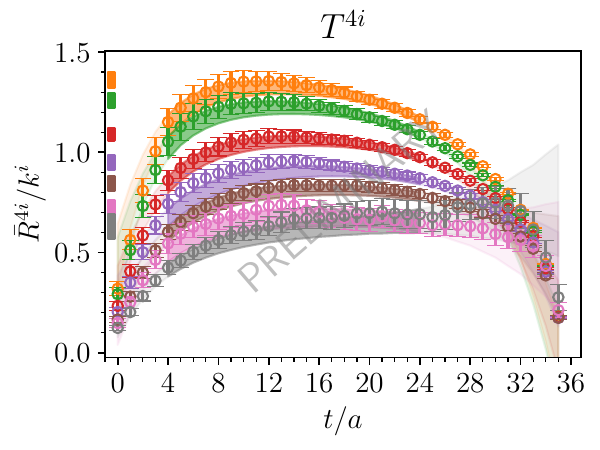}
    \caption{$B \to K$}
    \label{fig:ratio_B2K_T14}
  \end{subfigure}
  
  \begin{subfigure}[t]{.495\linewidth}
    \includegraphics[width=\linewidth]{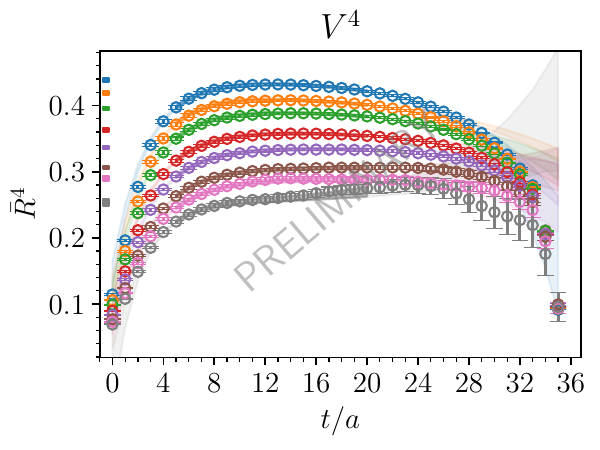}
    \caption{$B_s \to K$}
    \label{fig:ratio_Bs2K_V4}
  \end{subfigure}
  \hfill
  \begin{subfigure}[t]{.495\linewidth}
    \includegraphics[width=\linewidth]{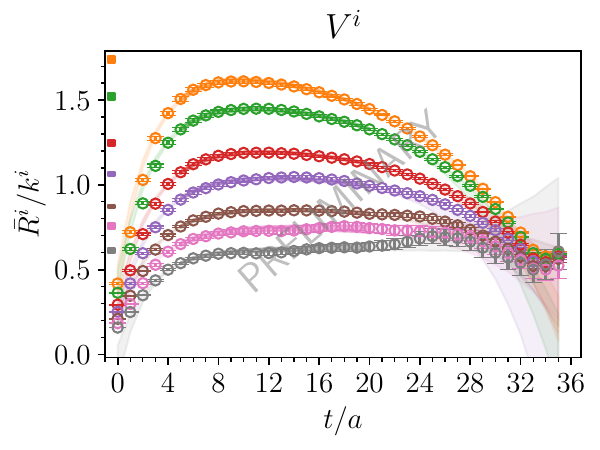}
    \caption{$B_s \to K$}
    \label{fig:ratio_Bs2K_V1}
  \end{subfigure}

  \caption{Averaged ratios $\overline{R}$ and their fit results (band). Here, $a \simeq 0.057 \fm$, and $m_l / m_s = \textrm{physical}$. The dark band region indicates the fit range. The leftmost colored box represents the fit result of the leading constant $F^{(0)}$ with the height as the uncertainty.}
  \label{fig:ratio}
\end{figure}
In Fig.~\ref{fig:ratio}, we present some examples of averaged ratios (divided by momentum $k^i$) and their preliminary fit results. In Fig.~\ref{fig:ratio_B2pi_T14} and \ref{fig:ratio_B2K_T14}, we show the ratios for the tensor current $T^{4i}$ of the $B \to \pi$ and $B \to K$ decays. In Fig.~\ref{fig:ratio_Bs2K_V4} and \ref{fig:ratio_Bs2K_V1}, we show the ratios for the vector currents $V^4$ and $V^i$ of the $B_s \to K$ decay. The leftmost colored boxes in the plots represent the fit results of the leading constant $F^{(0)}$, which corresponds to the form factors with normalization. For a given decay, fit ranges are chosen to be similar in physical units across ensembles.

\begin{figure}[tb]
  \centering
  \begin{subfigure}[t]{\linewidth}
    \includegraphics[width=\linewidth]{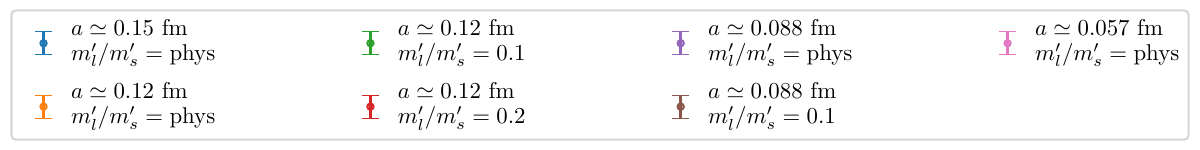}
  \end{subfigure}
  
  \begin{subfigure}[t]{.325\linewidth}
    \includegraphics[width=\linewidth]{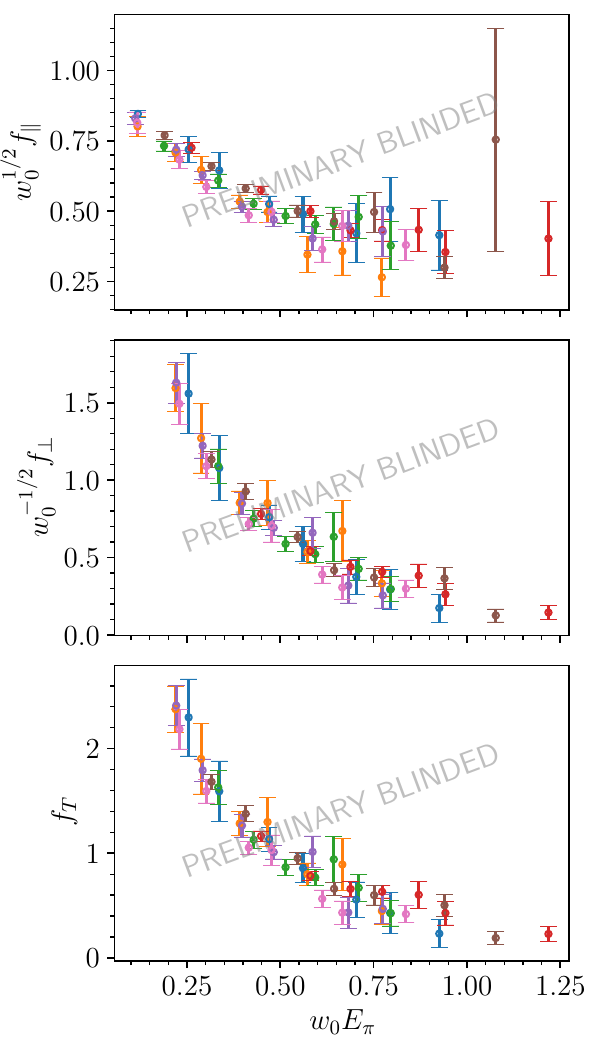}
    \caption{$B \to \pi$}
    \label{fig:ff_B2pi}
  \end{subfigure}
  \begin{subfigure}[t]{.325\linewidth}
    \includegraphics[width=\linewidth]{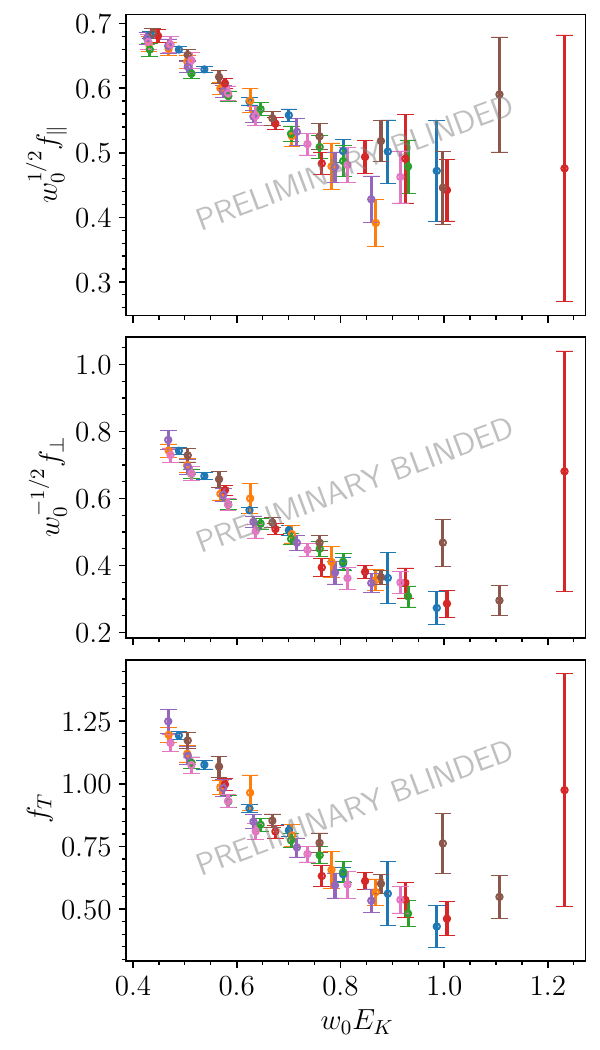}
    \caption{$B \to K$}
    \label{fig:ff_B2K}
  \end{subfigure}
  \begin{subfigure}[t]{.325\linewidth}
    \includegraphics[width=\linewidth]{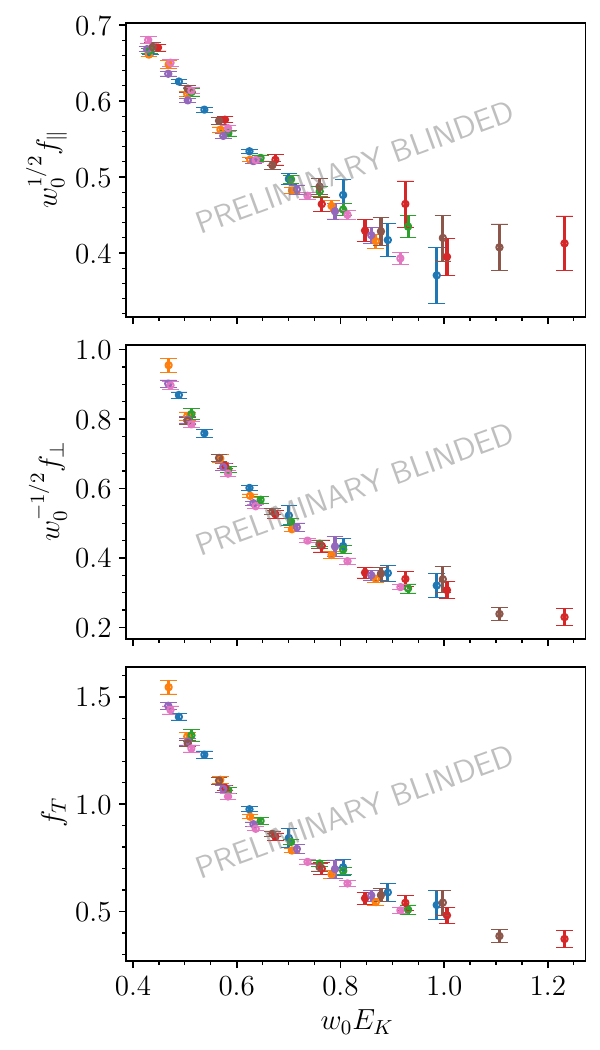}
    \caption{$B_s \to K$}
    \label{fig:ff_Bs2K}
  \end{subfigure}

  \caption{Form factors $f_\parallel$, $f_\perp$, and $f_T$ as a function of the recoil energy $\omega_0 E_L$.}
  \label{fig:ff}
\end{figure}
In Fig.~\ref{fig:ff}, we present some preliminary results for the lattice form factors $f_\parallel$, $f_\perp$, and $f_T$ as a function of the recoil energy $\omega_0 E_L$ for $L=\pi$, $K$. We use a mostly nonperturbative matching
\begin{equation}
  Z_J = \rho_J \sqrt{ Z_{V_{bb}^4} Z_{V_{qq}^4} } \,,
\end{equation}
where $q = l$ ($s$) for $L = \pi$ ($K$) \cite{Lepage:1992xa, El-Khadra:2001wco}. However, in this work, only the flavor-conserving renormalization factors $\sqrt{ Z_{V_{bb}^4} Z_{V_{qq}^4} }$ are applied, while the matching factor $\rho_J$ is not yet applied, so that we introduce an effective blinding of around $5\%$ into the analysis procedure, as in Refs.~\cite{FermilabLattice:2015mwy, Bailey:2015dka}.

\section{Summary and outlook}
\label{sec:conc}
We have calculated the complete set of lattice form factors $f_\parallel$, $f_\perp$, and $f_T$ for the $B \to \pi$, $B \to K$, and $B_s \to K$ decays on the (2+1+1)-flavor MILC HISQ gauge ensembles and presented some preliminary results. The HISQ action is used for the sea and light valence quarks, while the clover action in the Fermilab interpretation is used for the b quark.

The lattice form factors will be extrapolated to the continuum by means of heavy-meson rooted-staggered chiral perturbation theory (HMrS$\chi$PT), as in Ref.~\cite{FermilabLattice:2019ikx}. We will then extrapolate them to the full kinematic range accessible in the experiments by the model-independent $z$ expansion \cite{Boyd:1994tt} using the BCL parametrization \cite{Bourrely:2008za}. Finally, the form factors will be unblinded and used to compute the relevant decay rates or $|V_{ub}|$.
We also plan to combine our result with the collaboration's result for $B\to D^{(*)}$ form factors on the same ensembles to obtain the ratio $|V_{ub}| / |V_{cb}|$.

\acknowledgments{
  This work was supported in part 
  by the U.S. Department of Energy, Office of Science under grant Contract Numbers DE-SC0015655 (A.L., A.X.K.), DE-SC0010120 (S.G.), DE-SC0011090 (W.J.), DE-SC0021006 (W.J.);
  by the Simons Foundation under their Simons Fellows in Theoretical Physics program (A.X.K.);
  by the U.S. National Science Foundation under Grants No.~PHY17-19626 and PHY20-13064 (C.D., A.V.);
  by MCIN/AEI/10.13039/501100011033/FEDER, UE under Grants No.~PID2019-106087GB-C21 and PID2022-140440NB-C21 (E.G.);
  by the Junta de Andalucía (Spain) under Grant No.~FQM-101 (E.G.);
  and by AEI (Spain) under Grant No.~RYC2020-030244-I / AEI / 10.13039/501100011033 (A.V.).
  
  Computations for this work were carried out with resources provided by the USQCD Collaboration; by the ALCF and NERSC, which are funded by the U.S. Department of Energy; and by NCAR, NCSA, NICS, TACC, and Blue Waters, which are funded through the U.S. National Science Foundation. This research is part of the Blue Waters sustained-petascale computing project, which is supported by the National Science Foundation (awards OCI-0725070 and ACI-1238993) and the state of Illinois. Blue Waters is a joint effort of the University of Illinois at Urbana–Champaign and its National Center for Supercomputing Applications. Fermilab is operated by Fermi Research Alliance, LLC under Contract No.~DE-AC02-07CH11359 with the United States Department of Energy, Office of Science, Office of High Energy Physics.  
}

\nocite{*}
\bibliographystyle{JHEP-hc}
\bibliography{ref}

\providecommand{\href}[2]{#2}\begingroup\raggedright\begin{thebibliography}{10}

\bibitem{Gelzer:2017edb}
Z.~Gelzer et~al., \
  \href{https://doi.org/10.1051/epjconf/201817513024}{\emph{EPJ Web Conf.}
  {\bfseries 175} (2018) 13024}
  [\href{https://arxiv.org/abs/1710.09442}{{\ttfamily 1710.09442}}].

\bibitem{FermilabLattice:2019qpy}
{\scshape Fermilab Lattice, MILC} collaboration, \
  \href{https://doi.org/10.22323/1.363.0236}{\emph{PoS} {\bfseries LATTICE2019}
  (2019) 236} [\href{https://arxiv.org/abs/1912.13358}{{\ttfamily
  1912.13358}}].

\bibitem{Follana:2006rc}
{\scshape HPQCD, UKQCD} collaboration, \
  \href{https://doi.org/10.1103/PhysRevD.75.054502}{\emph{Phys. Rev. D}
  {\bfseries 75} (2007) 054502}
  [\href{https://arxiv.org/abs/hep-lat/0610092}{{\ttfamily hep-lat/0610092}}].

\bibitem{El-Khadra:1996wdx}
A.X.~El-Khadra, A.S.~Kronfeld and P.B.~Mackenzie, \
  \href{https://doi.org/10.1103/PhysRevD.55.3933}{\emph{Phys. Rev. D}
  {\bfseries 55} (1997) 3933}
  [\href{https://arxiv.org/abs/hep-lat/9604004}{{\ttfamily hep-lat/9604004}}].

\bibitem{MILC:2012znn}
{\scshape MILC} collaboration, \
  \href{https://doi.org/10.1103/PhysRevD.87.054505}{\emph{Phys. Rev. D}
  {\bfseries 87} (2013) 054505}
  [\href{https://arxiv.org/abs/1212.4768}{{\ttfamily 1212.4768}}].

\bibitem{Richardson:1978bt}
J.L.~Richardson, \
  \href{https://doi.org/10.1016/0370-2693(79)90753-6}{\emph{Phys. Lett. B}
  {\bfseries 82} (1979) 272}.

\bibitem{Bailey:2008wp}
J.A.~Bailey et~al., \
  \href{https://doi.org/10.1103/PhysRevD.79.054507}{\emph{Phys. Rev. D}
  {\bfseries 79} (2009) 054507}
  [\href{https://arxiv.org/abs/0811.3640}{{\ttfamily 0811.3640}}].

\bibitem{Lepage:1992xa}
G.P.~Lepage and P.B.~Mackenzie, \
  \href{https://doi.org/10.1103/PhysRevD.48.2250}{\emph{Phys. Rev. D}
  {\bfseries 48} (1993) 2250}
  [\href{https://arxiv.org/abs/hep-lat/9209022}{{\ttfamily hep-lat/9209022}}].

\bibitem{El-Khadra:2001wco}
A.X.~El-Khadra, A.S.~Kronfeld, P.B.~Mackenzie, S.M.~Ryan and J.N.~Simone, \
  \href{https://doi.org/10.1103/PhysRevD.64.014502}{\emph{Phys. Rev. D}
  {\bfseries 64} (2001) 014502}
  [\href{https://arxiv.org/abs/hep-ph/0101023}{{\ttfamily hep-ph/0101023}}].

\bibitem{FermilabLattice:2015mwy}
{\scshape Fermilab Lattice, MILC} collaboration, \
  \href{https://doi.org/10.1103/PhysRevD.92.014024}{\emph{Phys. Rev. D}
  {\bfseries 92} (2015) 014024}
  [\href{https://arxiv.org/abs/1503.07839}{{\ttfamily 1503.07839}}].

\bibitem{Bailey:2015dka}
J.A.~Bailey et~al., \
  \href{https://doi.org/10.1103/PhysRevD.93.025026}{\emph{Phys. Rev. D}
  {\bfseries 93} (2016) 025026}
  [\href{https://arxiv.org/abs/1509.06235}{{\ttfamily 1509.06235}}].

\bibitem{FermilabLattice:2019ikx}
{\scshape Fermilab Lattice, MILC} collaboration, \
  \href{https://doi.org/10.1103/PhysRevD.100.034501}{\emph{Phys. Rev. D}
  {\bfseries 100} (2019) 034501}
  [\href{https://arxiv.org/abs/1901.02561}{{\ttfamily 1901.02561}}].

\bibitem{Boyd:1994tt}
C.G.~Boyd, B.~Grinstein and R.F.~Lebed, \
  \href{https://doi.org/10.1103/PhysRevLett.74.4603}{\emph{Phys. Rev. Lett.}
  {\bfseries 74} (1995) 4603}
  [\href{https://arxiv.org/abs/hep-ph/9412324}{{\ttfamily hep-ph/9412324}}].

\bibitem{Bourrely:2008za}
C.~Bourrely, I.~Caprini and L.~Lellouch, \
  \href{https://doi.org/10.1103/PhysRevD.82.099902}{\emph{Phys. Rev. D}
  {\bfseries 79} (2009) 013008}
  [\href{https://arxiv.org/abs/0807.2722}{{\ttfamily 0807.2722}}].

\end{thebibliography}\endgroup

\end{document}